\documentclass{article}
\usepackage{spconf,amsmath,graphicx,amssymb}
\usepackage{subfigure}

\newtheorem{theorem}{Theorem}[section]

\title{Low-complexity optimization for Two-Dimensional Direction-\\ of-arrival Estimation via Decoupled Atomic Norm Minimization}
\name{Zhi Tian$^{\star}$, Zhe Zhang$^{\star \dagger}$, Yue Wang$^{\star}$\thanks{This work is partly supported by the US NSF grants \#CCF-1527396, \#ECCS-1546604, \#AST-1547329 and \#AST-1443858.}} 

\address{\normalsize $^{\star}$ Electrical and Computer Engineering Department,	George Mason University, Fairfax, VA 22030, USA \\
	\normalsize $^{\dagger}$ School of Engineering  and Applied Science, George Washington University, Washington, DC 20052, USA
	}
\begin{document}
%
\maketitle
\begin{abstract}
This paper presents an efficient optimization technique for super-resolution two-dimensional (2D) direction of arrival (DOA) estimation by introducing a new formulation of atomic norm minimization (ANM). ANM allows gridless angle estimation for correlated sources even when the number of snapshots is far less than the antenna size, yet it incurs huge computational cost in 2D processing.  
This paper introduces a novel formulation of ANM via semi-definite programming, which expresses the original high-dimensional problem by two decoupled Toeplitz matrices in one dimension, followed by 1D angle estimation with automatic angle pairing. Compared with the state-of-the-art 2D ANM, the proposed technique reduces the computational complexity by several orders of magnitude with respect to the antenna size, while retaining the benefits of ANM in terms of super-resolution performance with use of a small number of measurements, and robustness to source correlation and noise. The complexity benefits are particularly attractive for large-scale antenna systems such as massive MIMO and radio astronomy. 

\end{abstract}
\begin{keywords}
Two-dimensional DOA, atomic norm minimization, semi-definite programming, decoupled Toeplitz 
\end{keywords}
\section{Introduction}
\label{sec:intro}


The problem of 2D DOA estimation, as an instantiation of multivariate spectral analysis, arises in many applications such as azimuth-elevation angle estimation using 2D arrays and transceiver design in MIMO communications. Despite of the large body of literature \cite{hlv}, existing techniques can be quite complex for implementation in emerging large-scale antenna systems such as massive MIMO, where super-resolution 2D angle estimation need to be performed with low computing time from a small number of measurements. 

Subspace methods such as MUSIC and ESPRIT are popular for super-resolution 2D DOA estimation \cite{haardt19952d, roy1989esprit, hua1993pencil}. However, they all hinge on the sample covariance, which requires the number of snapshots to be larger than the antenna size. Besides, they are sensitive to signal correlation and noise, and may fail for coherent sources \cite{hlv}. 

Advances in compressed sensing (CS) suggests to exploit source sparsity for frequency or angular estimation \cite{donoho2006, candes2006stable}. CS enables DOA estimation even from a single snapshot of  measurements, regardless of source correlation. However, the angular estimates are confined on a finite-resolution grid, and the accuracy is sensitive to off-grid source mismatch \cite{chi2011sensitivity}. 

Recently, a new line of gridless CS for spectral analysis is developed via atomic norm minimization (ANM) in the form of semi-definite programming (SDP) \cite{tang2013compressed, candes2014towards}. It is a structure-based optimization approach where the Vandermonde structure of the array manifold is captured in the SDP via a Toeplitz matrix. It has been extended to the 2D case by exploiting  a two-level Toeplitz structure \cite{chi2015compressive, yang2016vandermonde}, which enjoys the benefits of the ANM approach in terms of  super-resolution from single-snapshot measurements, and resilience to source correlation. 
However, the computational load is heavy, which becomes near intractable for large-scale antenna systems. 

This paper presents a new formulation of ANM by introducing a new atom set that naturally decouples a two-level Toeplitz matrix into two Toeplitz matrices in one dimension. Accordingly, a new SDP formulation is developed for the decoupled ANM (D-ANM), which has a much reduced problem size and hence markedly improved computational efficiency. The time complexity is several orders of magnitude lower than that based on two-level Toeplitz, while other benefits of ANM  are preserved in terms of accuracy, resolution and use of a small number of snapshots. Analytic proof and simulations are presented in the paper to validate the proposed D-ANM  for low-complexity 2D DOA estimation. 


 

\section{RELATION TO PRIOR WORK}
\label{sec:rel}

\vspace{-0.05in}
This work belongs to the ANM-based optimization approach to spectral estimation \cite{candes2014towards, tang2013compressed, bhaskar2013atomic, yang2015gridless}, which has emerged as an effective alternative to traditional statistics-based approaches when the number of measurements is small. While SDP formulations for ANM are mostly done for 1D spectral estimation, this work is closely related to the recent 2D results in \cite{chi2015compressive, yang2016vandermonde}. Therein, the 2D array manifold matrix is vectorized into a long vector, whose structure is expressed in a SDP formula through a two-level Toeplitz matrix, followed by two-level Vandermonde decomposition.  For an $N\times M$ rectangular antenna array, the size of the SDP is $(NM+1)$ \cite{chi2015compressive, yang2016vandermonde}. This work presents a new SDP formula for decoupled ANM (D-ANM) with a small size of $(N+M)$, which drastically reduces the run time. The D-ANM not only avoids the cumbersome vectorization step in \cite{chi2015compressive, yang2016vandermonde}, but also enables the use of simple one-level Vandermonde decomposition and automatic angle pairing. Hence, both the SDP formulation for ANM and the ensuing Vandermonde decomposition for 2D DOA estimation are different from those in \cite{chi2015compressive, yang2016vandermonde}. 
 
\section{Signal Model}
\label{sec:bak}

Consider $K$ far-field narrowband waveforms $\mathbf{s}^\star=(s_1^\star, \dots, s_K^\star)^\mathrm{T}$ that impinge on an $N\times M$ uniform rectangular array with $N$ and $M$ elements along $x$-direction and $y$-direction respectively. The corresponding DOAs are denoted by $\boldsymbol{\theta}_x^\star=(\theta_{x, 1}^\star, \dots, \theta_{x, K}^\star)$ and $\boldsymbol{\theta}_y^\star=(\theta_{y, 1}^\star, \dots, \theta_{y, K}^\star)$ respectively. The noise-free baseband model for the array output matrix is
\begin{equation}
\label{eq:2.1}  
\mathbf{X}(t) = \sum_{k=1}^{K} s_k^\star(t) \mathbf{a}_N(\theta_{x, k}^\star)\mathbf{a}_M^\mathrm{H}(\theta_{y, k}^\star)
\end{equation}
where $\mathbf{a}_N(\theta_x)$ of length $N$ is the 1D array response vector in $x$-direction with a Vandemonde structure along $\theta_x$ \cite{hlv}, and $\mathbf{a}_M(\theta_y)$ of length $M$ is similarly defined. 

For clear exposition, we consider a single snapshot, thus dropping time $t$. Letting $\mathbf{A}_N(\boldsymbol{\theta}_x^\star)=[\mathbf{a}_N(\theta_{x, 1}^\star), \dots, \mathbf{a}_N(\theta_{x, K}^\star)]$, $\mathbf{A}_M(\boldsymbol{\theta}_y^\star)=[\mathbf{a}_M(\theta_{y, 1}^\star), \dots, \mathbf{a}_M(\theta_{y, K}^\star)]$, and $\mathbf{S}^\star=\mathrm{diag}(\mathbf{s}^\star)$, (\ref{eq:2.1}) can be concisely written as 
\begin{equation}
	\label{eq:2.4}
	\mathbf{X}=\mathbf{A}_N(\boldsymbol{\theta}_x^\star)\mathbf{S}^\star\mathbf{A}_M^\mathrm{H}(\boldsymbol{\theta}_y^\star).
\end{equation}
The goal of 2D DOA estimation is to recover $\boldsymbol{\theta}_x^\star$ and $\boldsymbol{\theta}_y^\star$ from observations of $\mathbf{X}$.  Here we focus on the noise-free signal structure to formulate an optimization approach. 

\section{The Atomic Norm Approach}

This section reviews the atomic norm minimization (ANM) approach to 2D DOA estimation in \cite{chi2015compressive, yang2016vandermonde}.   
Using the Kronecker product $\otimes$,  the signal $\mathbf{X}$ can be vectorized as \cite{chi2015compressive}
\begin{equation}
	\label{eq:2.5}
		\mathbf{x}=\mathrm{vec}(\mathbf{X})=\sum_{k=1}^{K} s_k^\star \mathbf{a}_M^\ast({\theta}_{y, k}^\star)\otimes\mathbf{a}_N({\theta}_{x, k}^\star)=\sum_{k=1}^{K} s_k \mathbf{a}(\boldsymbol{\theta}_k^\star)
\end{equation} 
where $\boldsymbol{\theta}=(\theta_{x}, \theta_{y})$, and $\mathbf{a}(\boldsymbol{\theta})=\mathbf{a}_M^\ast({\theta}_{y})\otimes\mathbf{a}_N(\theta_{x})$ is an extended array response vector of length $NM$. 

The atom set $\mathcal{A}_V$ is defined as
\begin{equation}
\label{eq:2.atom}
	\mathcal{A}_V=\{\mathbf{a}(\boldsymbol{\theta}), \quad \boldsymbol{\theta}\in[0, 2\pi]\times[0, 2\pi]\}.
\end{equation}

Let  $\mathbf{T}_\mathrm{2D}(\mathbf{u})$, defined by its first row $\mathbf{u}$ of length $NM$, denote a  two-level Hermitian Toeplitz matrix constructed from the two-level Vandemonde structure of $\mathbf{a}(\boldsymbol{\theta})$  \cite{chi2015compressive}. 
Then the atomic norm of $\mathbf{x}$ can be calculated via SDP: 
\begin{eqnarray}
			\|\mathbf{x}\|_{\mathcal{A}_V}&\hspace{-0.08in}=&\hspace{-0.08in}\inf \left\{\sum_k |s_k| \left| \mathbf{x}=\sum_k s_k \mathbf{a}(\boldsymbol{\theta}_k), \ \ \mathbf{a}(\boldsymbol{\theta})\in\mathcal{A}_V \!\! \right. \right\} \nonumber \\
			&\hspace{-0.08in}=&\hspace{-0.08in}\min_{\mathbf{u}, v} \ \left\{\frac{1}{2}\left(v+\mathrm{trace}\big(\mathbf{T}_\mathrm{2D}(\mathbf{u})\big)\right)\right\} \nonumber \\
			&& \mathrm{s.t.}\quad \left(\begin{array}{cc}
			v & \mathbf{x}^\mathrm{H} \\
			\mathbf{x} & \mathbf{T}_\mathrm{2D}(\mathbf{u})
			\end{array}\right)\succeq \mathbf{0}. \label{eq:2.6}
\end{eqnarray}

It has been shown  that if $\mathbf{x}$ is composed of only a few adequately-separated sources, then $(\boldsymbol{\theta}_x^\star, \boldsymbol{\theta}_y^\star)$ can be recovered by computing $\|\mathbf{x}\|_{\mathcal{A}_V}$ from (noiseless, noisy or partial) observations of $\mathbf{x}$ \cite{chi2015compressive}. The SDP in (\ref{eq:2.6}) results in the two-level Toeplitz matrix $\mathbf{T}_\mathrm{2D}(\mathbf{u})$, which contains the angular information from both dimensions and can be processed via two-level Vandemonde decomposition to yield $(\boldsymbol{\theta}_x^\star, \boldsymbol{\theta}_y^\star)$  \cite{yang2016vandermonde}.

The main issue of the vectorization-based ANM in (\ref{eq:2.6}) is its high complexity. Due to the vectorization in (\ref{eq:2.5}), the matrix size in the SDP constraint is $(NM+1) \times (NM+1)$, which incurs high complexity in both computation and memory as $N$ and $M$ become large. We tried simulations on a PC for $N=M=32$, in which case the SDP calculation could not finish in two days. For large-scale antenna systems, an efficient implementation of the ANM principle is motivated.

\section{2D DOA estimation via Decoupled ANM}
\label{sec:main}

This section presents the main results, namely a decoupled ANM formulation for efficient 2D DOA estimation. 

\subsection{Decoupled ANM and its SDP reformulation}

Recall the signal model  in (\ref{eq:2.1}) and (\ref{eq:2.4}). Alternative to the vectorized atom set in (\ref{eq:2.atom}), we adopt a new atom set $\mathcal{A}_M$ as
\begin{equation}
	\label{eq:3.1}
	\begin{split}
		\mathcal{A}_M&=\left\{\mathbf{a}_N(\theta_x)\mathbf{a}_M^\mathrm{H}(\theta_y), \quad \theta_x\in[0, 2\pi], \theta_y\in[0, 2\pi]\right\} \\
			&=\left\{\mathbf{A}(\boldsymbol{\theta}), \quad \boldsymbol{\theta}\in[0, 2\pi]\times[0, 2\pi]\right\}.
	\end{split}
\end{equation}

Our approach to find $(\boldsymbol{\theta}_x^\star, \boldsymbol{\theta}_y^\star)$ from $\mathbf{X}$ is to find the following atomic norm:
\begin{equation}
	\label{eq:3.2}
 \|\mathbf{X}\|_{\mathcal{A}_M}\!=\inf \left\{\!\sum_k |s_k| \left| \mathbf{X}=\!\sum_k s_k \mathbf{A}(\boldsymbol{\theta}_k), \ \ \mathbf{A}(\boldsymbol{\theta})\in\mathcal{A}_M \!\! \right. \right\}.
\end{equation}

This is an infinite programming problem over all feasible $\boldsymbol{\theta}$. 
By reformulating (\ref{eq:3.2}) via SDP, our main results follow. 

\begin{theorem}
	\label{th:1}
	Consider an $N\times M$ data matrix $\mathbf{X}$ given by
	\begin{equation}
		\label{eq:u2}
		\mathbf{X}=\textstyle \sum_{k=1}^{K} s_k^\star \mathbf{A}(\boldsymbol{\theta}_k^\star). 
	\end{equation}
Define the minimal angle distances as $\Delta_{\min, x}=\min_{i\neq j} |\sin \theta_{x, i}^\star-\sin \theta_{x, j}^\star|$ and $\Delta_{\min, y}=\min_{i\neq j} |\sin \theta_{y, i}^\star-\sin \theta_{y, j}^\star|$, 
which are wrapped distances on the unit circle. If they satisfy
	\begin{equation}
		\Delta_{\min, x}\geq\frac{1.19}{\lfloor(N-1)/4\rfloor}, \mathrm{~~~}\Delta_{\min, y}\geq\frac{1.19}{\lfloor(M-1)/4\rfloor}, 
	\end{equation}
then it is guaranteed that (\ref{eq:u2}) is the optimal solution to (\ref{eq:3.2}). Further, it can be efficiently computed via the following SDP:	
	\begin{eqnarray}
		\|\mathbf{X}\|_{\mathcal{A}_M}&\hspace{-0.15in}=&\hspace{-0.15in}\min_{\mathbf{u}_x, \mathbf{u}_y}\! \!\left\{\! \frac{1}{2\sqrt{NM}}\bigg(\!\mathrm{trace}\big(\mathbf{T}(\mathbf{u}_x)\big)\!+\!\mathrm{trace}\big(\mathbf{T}(\mathbf{u}_y)\big)\!\!\bigg)\!\right\} \nonumber \\
		&&\mathrm{s.t.}\quad \left(\begin{array}{cc}
		\mathbf{T}(\mathbf{u}_y) & \mathbf{X}^\mathrm{H} \\
		\mathbf{X} & \mathbf{T}(\mathbf{u}_x)
		\end{array}\right)\succeq \mathbf{0}\label{eq:3.3}
	\end{eqnarray}
	where $\mathbf{T}(\mathbf{u}_x)$ and $\mathbf{T}(\mathbf{u}_y)$ are one-level Hermitian Toeplitz matrices defined by the first rows $\mathbf{u}_x$ and $\mathbf{u}_y$ respectively.
\end{theorem}


\noindent {\bf \em Remark 1: Angular information.} At the optimal $\hat{\mathbf{u}}_x$, the $N\times N$ Toeplitz matrix $\mathbf{T}(\hat{\mathbf{u}}_x)$ reveals $\boldsymbol{\theta}_x^\star$ via  
\begin{equation} \label{eq:Tx} 
\mathbf{T}(\hat{\mathbf{u}}_x)=\mathbf{A}_N(\boldsymbol{\theta}_x^\star) \mathbf{D}_x \mathbf{A}_N^{\mathrm{H}}(\boldsymbol{\theta}_x^\star), \quad \mathbf{D}_x \succeq \mathbf{0} \mathrm{~is~diagonal}. 
 \end{equation}
Similarly,  $\boldsymbol{\theta}_y^\star$ is  coded in the $M\times M$ matrix $\mathbf{T}(\hat{\mathbf{u}}_y)$. 
Hence, after the SDP in (\ref{eq:3.3}), 2D DOA information can be acquired via Vandemonde decomposition on these one-level Toeplitz matrices, which is much simpler than the two-level Vandemonde decomposition needed in \cite{chi2015compressive, yang2016vandermonde}, and there are many mature techniques such as 
subspace methods, matrix pencil \cite{yang2016vandermonde} and Prony's method \cite{yang2015gridless}.

\noindent{\bf \em Remark 2: Decoupling.} The main benefit of the new result (\ref{eq:3.3}) is its low complexity via a decoupled formulation for ANM.  Instead of coupling the 2D DOA  via vectorization to form a constraint of size $(NM+1)\times (NM+1)$ in (\ref{eq:2.6}), (\ref{eq:3.3}) decouples the angular information into two one-level Toeplitz matrices, which markedly reduces the constraint size in SDP to $(N+M)\times(N+M)$. 

\subsection{Sketch of proof}
\label{sec:prf}

The proof for Theorem \ref{th:1} will be detailed in a journal version. It is outlined next under the page limit. 

\subsubsection{Uniqueness of atomic norm via dual polynomial}

Define the dual norm of $\|\cdot\|_{\mathcal{A}_M}$ for $\forall\mathbf{Q}\in\mathbb{C}^{N\times M}$ as 
\begin{equation}
	\|\mathbf{Q}\|_{\mathcal{A}_M}^\ast=\sup_{\|\mathbf{X}\|_{\mathcal{A}_M}\leq 1} \langle\mathbf{Q}, \mathbf{X}\rangle_\mathcal{R}
\end{equation}
where $\langle\cdot, \cdot\rangle$ denotes the Frobenius inner product, and $\langle\cdot, \cdot\rangle_\mathcal{R}=\Re\left\{\langle\cdot, \cdot\rangle\right\}$ keeps the real part.

Following standard Lagrangian analysis as in \cite{tang2013compressed, boyd2004convex}, 
one can show that if there exists a dual polynomial
\begin{equation}
	Q(\boldsymbol{\theta}) =\langle \mathbf{Q}, \mathbf{A}(\boldsymbol{\theta})\rangle, \quad  \mathbf{Q}\in\mathbb{C}^{N\times M}, \ \mathbf{A}(\boldsymbol{\theta})\in\mathcal{A}_M
\end{equation}
satisfying the conditions (bounded interpolation property)
	\begin{equation}
	\label{eq:c1+2}
\begin{split}
		Q(\boldsymbol{\theta}_k^\star)&=\mathrm{sign}(s_k^\star), \quad \forall \boldsymbol{\theta}_k^\star\in\Omega;
\\
		|Q(\boldsymbol{\theta})|&<1, \quad \quad \quad \quad \forall \boldsymbol{\theta}\notin\Omega,
\end{split}
	\end{equation}
then it is guaranteed that the optimal solution to (\ref{eq:3.2}) is unique, where $\Omega=\{\boldsymbol{\theta}_1^\star, \dots, \boldsymbol{\theta}_K^\star\}$ collects all supports of $\mathbf{X}$. 

The rest of proof is to find a dual polynomial that satisfies the above conditions, following a similar procedure as in \cite{chi2015compressive}.

\subsubsection{Equivalence of atomic norm to SDP with decoupling}

Denote the optimal solution of (\ref{eq:3.3}) as $\mathrm{SDP}(\mathbf{X})$. On one hand, for arbitrary atomic decomposition of $\mathbf{X}=\sum_k s_k \mathbf{A}(\boldsymbol{\theta}_k)$, it is easy to verify the semi-definite contraint in (\ref{eq:3.3}) by letting $\mathbf{T}(\mathbf{u}_x)=\sum_k |s_k| \sqrt{\frac{M}{N}}   \mathbf{a}_N(\theta_{x, k}) \mathbf{a}_N^\mathrm{H}(\theta_{x, k})$ and $\mathbf{T}(\mathbf{u}_y)=\sum_k |s_k| \sqrt{\frac{N}{M}} \mathbf{a}_M(\theta_{y, k}) \mathbf{a}_M^\mathrm{H}(\theta_{y, k})$; hence, $\mathrm{SDP}(\mathbf{X})\leq\|\mathbf{X}\|_{\mathcal{A}_M}$.
	
On the other hand, from the semi-definite condition, $\mathbf{X}$ lies in the column space of $\mathbf{T}(\mathbf{u}_x)$ and row space of $\mathbf{T}(\mathbf{u}_y)$. Similar to \cite{yang2014exact}, using Schur completion lemma and geometry averaging inequality, one can verify $\mathrm{SDP}(\mathbf{X})\geq\|\mathbf{X}\|_{\mathcal{A}_M}$.

\subsection{2D DOA Estimation based on decoupled ANM} 

In practice, $\mathbf{X}$ is observed implicitly. We consider a linear observation model in the presence of an additive noise $\mathbf{W}$: 
\begin{equation}
\label{eq:5.3.1}
	\mathbf{Y}= \mathcal{L}(\mathbf{X}) + \mathbf{W}
\end{equation}
where $\mathcal{L}(\cdot)$ represents linear mixing, with possible missing entries and down-sampling. Given $\mathbf{Y}$, and exploiting the signal structure (\ref{eq:3.3}),  $\mathbf{X}$ can be estimated via regularization 
\begin{equation}
	\min_\mathbf{X} \quad \lambda\|\mathbf{X}\|_{\mathcal{A}_M}+\|\mathbf{Y}-\mathcal{L}(\mathbf{X})\|_\mathrm{F}^2
\end{equation}
where $\lambda\geq 0$ is a weighting scalar. To find the 2D DOA, it is adequate to find the two desired Toeplitz matrices via
\begin{equation}
		\label{eq:5.3.3}
		\begin{split}
		\min_{\mathbf{u}_x, \mathbf{u}_y, \mathbf{X}} &\quad \frac{\lambda}{2\sqrt{NM}}\bigg(\!\mathrm{trace}\big(\mathbf{T}(\mathbf{u}_x)\big)+\mathrm{trace}\big(\mathbf{T}(\mathbf{u}_y)\big)\!\!\bigg)  \\
		 & \quad +\|\mathbf{Y}-\mathcal{L}(\mathbf{X})\|_\mathrm{F}^2\\
		\mathrm{s.t.} & \quad \left(\begin{array}{cc}
		\mathbf{T}(\mathbf{u}_y) & \mathbf{X}^\mathrm{H} \\
		\mathbf{X} & \mathbf{T}(\mathbf{u}_x)
		\end{array}\right)\succeq \mathbf{0}.
		\end{split}
	\end{equation}
As mentioned in {\bf \em Remark 1}, mature 1D DOA estimators can be employed to obtain the estimates $\hat{\boldsymbol{\theta}}_{x}$ and $\hat{\boldsymbol{\theta}}_y$ from the optimal  $\mathbf{T}(\hat{\mathbf{u}}_x)$ and $\mathbf{T}(\hat{\mathbf{u}}_y)$, in a decoupled manner. 

Like all 2D DOA estimators, a  pairing step is critical to find $(\hat{\theta}_{x, k}, \hat{\theta}_{y, k})$ pairs, $\forall k$. Since we have $\hat{\mathbf{X}}$ at hand via (\ref{eq:5.3.3}), we develop a simple angle pairing technique as follows. 
\begin{enumerate}
\itemsep -0.03in
	\item Construct the array response matrix $\mathbf{A}_N(\hat{\boldsymbol{\theta}}_x)$ from $\hat{\boldsymbol{\theta}}_x$.
	\item Compute $\mathbf{V}_M=\mathbf{D}_x^{-1}\mathbf{A}_N^\dagger(\hat{\boldsymbol{\theta}}_x)\hat{\mathbf{X}}$, where $(\cdot)^\dagger$ denotes pseudo-inverse and $\mathbf{D}_x$ is the diagonal matrix in (\ref{eq:Tx}) obtained from Vandermonde decomposition of $\mathbf{T}(\hat{\mathbf{u}}_x)$. In the noise-free case, $\mathbf{V}_M^\mathrm{H} = [\mathbf{v}_1, \ldots, \mathbf{v}_K]$ is the same as $\mathbf{A}_M(\boldsymbol{\theta}_y^\star)$ up to phase ambiguity and global scaling. 
	\item Pair up $\hat{\theta}_{y,k}$ with $\hat{\theta}_{x,k_y}$ via maximum correlation: 
	\[ k_y = \arg \max_{j\in [1, K]}  \left| \left\langle \mathbf{a}_M(\hat{\theta}_{y,k}), \mathbf{v}_j \right \rangle \right|, \quad k = 1, \ldots, K. \]
\end{enumerate}

\section{Performance Evaluation}
\label{sec:sim}

\subsection{Complexity analysis}


As mentioned in {\bf Remark 2}, 
in the {\em  vectorized ANM} method \cite{chi2015compressive}, the semi-definite constraint in (\ref{eq:2.6}) is of size $(NM+1)\times(NM+1)$. As a result, the SDP step for recovering the two-level Toeplitz has time complexity $\mathcal{O}(N^{3.5}M^{3.5}\log(1/\epsilon))$, where $\epsilon$ is the desired recovery precision \cite{krishnan2005interior}; the ensuing step of two-level Vandermonde decomposition for angle estimation and pairing has time complexity $\mathcal{O}(N^2 M^2 K)$ \cite{yang2016vandermonde}. 

In contrast, our {\em decoupled ANM} formulation in (\ref{eq:3.3}) and (\ref{eq:5.3.3}) has a constraint of smaller size $(N+M)\times(N+M)$. The time complexity of the SDP step is $\mathcal{O}((N+M)^{3.5}\log(1/\epsilon))$, and that of the one-level Vandermonde decomposition and pairing step is $\mathcal{O}((N^2 +M^2) K)$. The overall complexity is reduced by an order of $\mathcal{O}(N^{3.5})$ for $N=M$.  

\begin{figure}[t]
	\centering
	\includegraphics[width=.93\columnwidth]{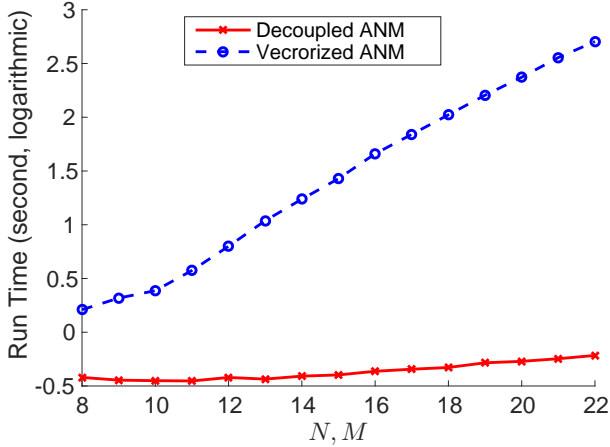}	
	\vspace{-0.1in}
	\caption{Computing complexity: run time versus $N$ ($N=M$).}
	\label{fig:3}
\end{figure}

Simulations of the run time are performed for a square array with $M=N$, and $K=4$ sources. The running speeds of these two methods are plotted on a logarithmic scale versus $N$ in Figure \ref{fig:3}. Our method exhibits a huge benefit in computational efficiency for large-scale arrays. When $M=N=22$, the running time of the vectorized ANM is 733.1842s, while that of the decoupled ANM is only 1.4997s.

\subsection{Recovery accuracy and noise performance}

Mont\`{e} Carlo simulations are carried out to evaluate the DOA estimation performance of both ANM-based 2D DOA methods, with $M=N=16$ and $K=4$.  Following SDP, the matrix pencil method is used to carry out both one- and two-level Vandermonde decomposition for DOA estimation \cite{yang2016vandermonde, hua1993pencil}.

\begin{figure}[t]
	\centering
		\includegraphics[width=.93\columnwidth]{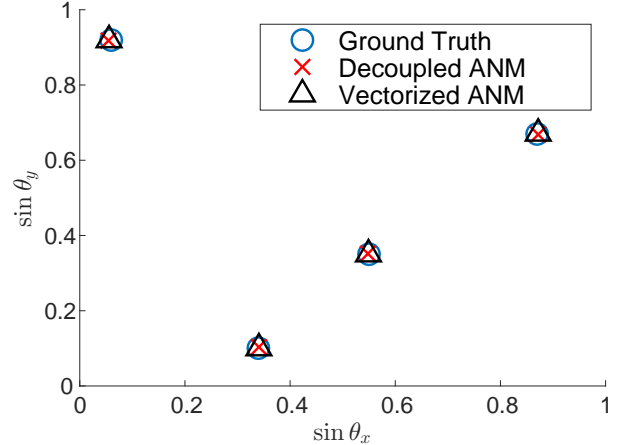}	
		\vspace{-0.1in}
	\caption{2D DOA (SNR = 20dB).}
	\label{fig:1}
\end{figure}

Figure \ref{fig:1} depicts the estimated 2D angles at a high signal to noise ratio (SNR) of 20dB. All source angles are accurately recovered, which confirms that both ANM methods correctly capture the inherent signal structures in their formulations. 

\begin{figure}[t]
	\centering
		\includegraphics[width=.93\columnwidth]{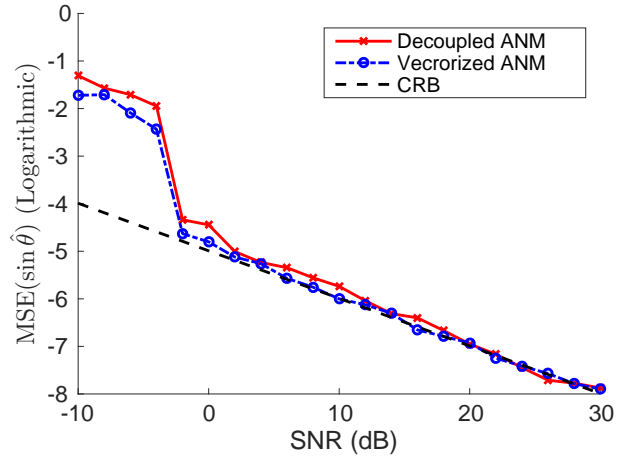}
		\vspace{-0.1in}
	\caption{Noise performance: MSE vs. SNR ($N=M=16$).}
	\label{fig:4}
\end{figure}

Figure \ref{fig:4} compares the average mean square error (MSE) of the estimates $\sin\boldsymbol{\hat{\theta}}$ versus SNR, with reference to the Cramer-Rao bound (CRB)  \cite{liu2007multidimensional}. The MSE performance of the proposed decoupled ANM is quite close to that of the vectorized ANM, both approaching the CRB for large SNR. 

\section{Conclusion}
\label{sec:con}

This work presents a novel decoupled ANM approach to 2D DOA estimation. By introducing a new atom set and decoupling the angular information onto lower dimension, we have reduced the computational load by several orders of magnitude in array size, while retaining the benefits of ANM in terms of gridless estimation, light  measurement requirements and robustness to signal correlation. Automatic angle pairing is also developed. The proposed low-complexity optimization technique can be extended to other high-dimensional spectral estimation problems as well. Future work includes performance analysis in the presence of data compression and noise.




\bibliographystyle{IEEEbib}
\bibliography{overview}

\end{document}